\documentclass[a4paper,11pt]{article}
\usepackage{jheppub}
\usepackage{braket}
\usepackage{mathtools}

\title{One-loop mass corrections of interacting string states}

\author[a,b]{Lorenzo Grimaldi}
\author[a]{Massimo Bianchi}
\author[c]{Maurizio Firrotta}

\affiliation[a]{Dipartimento di Fisica, Universit\`a di Roma Tor Vergata and INFN sezione di Roma Tor Vergata,\\
  Via della Ricerca Scientifica 1, 00133, Roma, Italy}

\affiliation[b]{`Enrico Fermi' Research Center,
  Via Panisperna 89A, 00184, Roma, Italy}

\affiliation[c]{Department of Physics and Center for Field Theory and Particle Physics, Fudan University\\
Shanghai 200438, China}

\emailAdd{lorenzo.grimaldi@roma2.infn.it}
\emailAdd{massimo.bianchi@roma2.infn.it}
\emailAdd{mfirrotta@fudan.edu.cn}

\abstract{The free string spectrum is highly degenerate, with a degeneracy that grows exponentially with the mass. Turning on a non-vanishing string coupling $g_s$ introduces interactions, rendering massive string states unstable and allowing them to decay into lower-mass states, with mixing constrained by Lorentz invariance. This behavior is expected already at one-loop level. The imaginary part of the one-loop mass correction is related to the width of the decay into two lower-mass states at tree level, whereas the real part is generally IR-divergent and needs regularization and renormalization.
The analysis simplifies for states in the first Regge trajectory.

In particular, we consider the one-loop mass corrections for these states in the NS-NS sector of Type-II string theories. We explicitly construct the related vertex operators and exploit the properties of elliptic functions in order to obtain a closed form expression for the integral over the insertion point. We further regularize the divergences of the integral over the modular parameter of the torus by means of the $i\varepsilon$-prescription in string theory.
Finally we extract numerical results for the mass correction up to level $N=4$.}

\begin{document}
\maketitle

\section{Motivations}
String theory was originally introduced to describe hadronic resonances, which are known to exhibit level repulsion: states with identical quantum numbers are shifted apart in mass and, after unfolding, their spectra follow the Wigner–Dyson distribution \cite{Dyson:1962es,Dyson:1962oir,Wigner:1967qdh} characteristic of the Gaussian Orthogonal Ensemble. Although string theory is now viewed as a candidate framework for all fundamental interactions, it remains important to ask whether level repulsion is an intrinsic feature of string theory itself. Initial results for the bosonic string have been obtained in \cite{Bianchi:2022mhs,Bianchi:2023uby}, and the supersymmetric case can be investigated along similar lines.

A further motivation for studying the string spectrum is its exponential degeneracy with mass \cite{Hagedorn:1965st}. This feature makes highly excited string states natural candidates for black-hole microstates in string theory, whose complexity is comparable (see e.g. \cite{Susskind:1993ws,Damour:1999aw,Horowitz_1997} and recent developments \cite{Firrotta:2024fvi}). This perspective motivates the study of one-loop mass corrections, which are technically challenging concerning both vertex operators and the arising of IR divergences \cite{Marcus:1988vs,Iengo:2002tf,Sen:2016gqt,Stieberger:2023nol}. Recent progress \cite{Manschot:2024prc} has enabled reliable computations for a special class of string states, which we investigate here.

\section{NS-NS spectrum and one-loop amplitudes}
We focus on the NS-NS sector of Type II string theories, for which the lowest states are massless, featuring the symmetric traceless graviton $g_{MN}$, the Kalb-Ramond 2-form field $B_{MN}$, and the dilaton $\phi$. We will always assume the level matching condition $N=\overline{N}$, where $N$ and $\overline{N}$ are positive integers. The mass of a state at level $N$ is given by $\alpha'M^2_N = 4(N-1)$. We will set $\alpha'=2$.

Mass corrections arise from one-loop contributions to the propagator and are therefore extracted from one-loop two-point amplitudes. The loop expansion of amplitudes is actually a genus expansion. At any given order in the string coupling $g_s$, the interaction is represented by two closed strings inserted on a worldsheet in the Euclidean signature, with a suitable genus. At one-loop level, such a genus-1 worldsheet is the torus, which is identified by a modular parameter $\tau=\tau_1+i\tau_2$ and parametrized by $z=x+\tau y$, with $x,\,y\in[0,\,1]$. The interaction at a given genus is encoded in the vertex operators ${\cal V}_i(z,\bar{z})$ that enter with each string insertion. The isometry group of the torus allows one insertion point to be fixed, whereas we must integrate over the remaining point. Furthermore, as a consequence of the modular invariance of the torus, we only have to consider the inequivalent tori. Hence, we must further integrate over the fundamental domain $\mathcal{F}:=\left\{\tau\in\mathbb{C}\,:\,-1/2\leq\tau_1\leq 1/2,\,\sqrt{1-\tau_1^2}\leq\tau_2<+\infty\right\}$ of modular parameters. The interested reader may refer to \cite{Kiritsis:2019npv} for a pedagogical introduction. 

Here we restrict our attention to the states of the leading Regge trajectory, namely the states for which the spin is maximal and equal to $2N$. This simplification enables us to provide a compact form of the corresponding vertex operators. Additionally, since this Lorentz representation is unique at each level, Lorentz invariance forbids any perturbative mixing of these states with other states at the same level,  thus diagonalizing the mixing matrix in this sector. Furthermore, notice that states at different mass levels cannot mix in perturbation theory.

\section{Mass corrections to the leading Regge trajectory}
The  vertex operator of the NS-NS states in the leading Regge trajectory in the $0$-picture\footnote{The vertex operator \eqref{open string vertex} is derived from the one in the (${-}1$)-picture ${\cal V}^{(-1)} = e^{-\varphi}\psi^{\mu_1}\left(\prod_{i=2}^N \partial X^{\mu_i}\right) e^{iPX}$ through the picture-changing operator $\Gamma_{+1} = e^{\varphi} \psi{\cdot} \partial X {+} ...$, with $\varphi$ the scalar bosonizing the super-ghost ($\beta, \gamma$) system \cite{Kiritsis:2019npv}.} is 
\begin{equation}
{\cal W} = \frac{1}{(N-1)!}H_{\mu_1\dots\mu_N\bar{\mu}_1\dots\bar{\mu}_N}{\cal V}^{\mu_1\dots\mu_N} {\overline{\cal V}}^{\bar{\mu}_1\dots\bar{\mu}_N},
\end{equation}
where $H_{\mu_1\dots\bar{\mu}_N}\sim\bigotimes_{i=1}^N\lambda_{\mu_i}\lambda_{\bar{\mu}_i}$ is a totally symmetric, transverse, traceless tensor,
\begin{flalign}
    \mathrlap{{\cal V}^{\mu_1\dots\mu_N} {=} [\partial X^{\mu_1} \partial X^{\mu_2} {+} {(}{N}{-}{1}{)} P{\cdot}\psi \psi^{\mu_1}\partial X^{\mu_2} {+}{(}{N}{-}{1}{)}{(}{N}{-}{2}{)} \psi^{\mu_1} \partial\psi^{\mu_2}] \left(\prod_{i=3}^N \partial X^{\mu_i}\right)e^{iPX}}&&
    \label{open string vertex}
\end{flalign}
and $\overline{{\cal{V}}}^{\bar{\mu}_1,\dots,\bar{\mu}_N}$ is the complex conjugate of \eqref{open string vertex}. The one-loop scattering amplitude for the mass corrections of the above massive states thus reads (up to an overall factor)
\begin{equation}
\mathcal{A}^{(N)}:=\sum_{s_1,s_2} c_{s_1}c_{s_2}\int_{\mathcal{F}}\frac{d^2\tau}{\tau_2^2}\,{\cal Z}_{(s_1,s_2)}\left<\int_{T^2}d^2z\,\mathcal{W}(z,\bar{z}) \int_{T^2}d^2w\,\mathcal{W}(w,\bar{w})\right>_{(s_1,s_2)},
\end{equation}
where $s_1$ and $s_2$ denote the spin structures on the torus $T^2$ and the Type II partition function reads
\begin{equation}
{\cal Z}_{(s_1,s_2)}:= \frac{\vartheta_{s_1}^4(0|\tau)\overline{\vartheta_{s_2}(0|\tau)}^4}{4\tau_2^{4}\eta(\tau)^{12}\overline{\eta(\tau)}^{12}},
\label{partition function}
\end{equation}
with $\eta$ the Dedekind function and $\vartheta_s$ the Jacobi theta functions ($s=1,\dots,4$) defined in, {\it e.g.}, \cite{Kiritsis:2019npv}.

We must now perform the Wick contractions between the bosonic and fermionic coordinates. In the holomorphic sector, the fermionic propagator (Szego kernel) is defined as
\begin{equation}
\begin{split}
\langle \psi(u_i) \psi(u_j) \rangle_{T^2}^{s}=\begin{cases} S_s(u_{ij}):= \frac{\vartheta_s(u_{ij}|\tau)}{\vartheta_1(u_{ij}|\tau)}  \frac{\partial_u\vartheta_1(0|\tau)}{\vartheta_s(0|\tau)} \,, \quad s=2,3,4  \\ S(u_{ij}):=  -\partial_{u_i}\langle X(u_i)X(u_j) \rangle_{T^2} \,, \,\,\,\, s=1\end{cases}
\end{split}
\end{equation}
 with $u_{ij}=u_i-u_j$, whereas the bosonic propagator (Bargmann kernel) is
\begin{equation}
\langle X(u_i,\bar{u}_i)X(u_j,\bar{u}_j) \rangle_{T^2}= {G}(u_{ij},\bar{u}_{ij}):=-\log \left| \frac{\vartheta_1(u_{ij} |\tau)}{\partial_u\vartheta_1(0|\tau)}   \right|^{2} + 2\pi\frac{({\rm Im}\,u_{ij})^2}{\tau_2}.
\end{equation}
The integration of the zero-mode of the bosonic coordinates results in momentum conservation among contracted vertex operators $P_1=-P_2=P$. Assuming the transversality relations $\lambda_k^{(1)} {\cdot} P_1 = 0= \lambda_k^{(2)} {\cdot} P_2$, this further produces $\lambda_k^{(1)} {\cdot} P_2 = 0= \lambda_k^{(2)} {\cdot} P_1$. Hence, all the contractions between $e^{\pm iPX}$ and $\partial X$ terms vanish \cite{Manes:1988gz}. The only bosonic contributions are thus ($z=u_1-u_2$)
\begin{equation}
\langle e^{iP_1X} e^{iP_2X} \rangle = e^{-P_1P_2 G(z,\bar{z})} = e^{-M^2 G(z,\bar{z})}  = \left| \frac{\vartheta_1(z|\tau)}{\partial_z\vartheta_1(0|\tau)} \right|^{4(N-1)} e^{-4\pi(N-1) \frac{({\rm Im}\,z)^2}{\tau_2}},
\end{equation}
\begin{equation}
 \langle \prod_{i=3}^N \partial X^{\mu_i}(u_1)  \prod_{j=3}^N \partial X^{\mu_j} (u_2)\rangle  = \left[\partial_z^2 G(z)\right]^{(N-2)}. 
\end{equation}
For each spin structure $s$, the fermionic contractions schematically amount to
\begin{equation}
\langle \prod_{i=1}^2(\partial_iX)^N\rangle_s {+} \langle \prod_{i=1}^2{P_i}{\cdot}{\psi} {\psi} (\partial_i X)^{N-1}\rangle_s  {+} \langle \prod_{i=1}^2{\psi} {\partial_i}{\psi} (\partial_i X)^{N-2}\rangle_s.
\end{equation}
Due to the Jacobi and Riemann identities, after summing over $s$, only the last term survives:
\begin{equation}
\langle \psi \partial\psi (\partial X)^{N-2}
\psi \partial\psi (\partial X)^{N-2} \rangle_s \sim  [\partial_z\vartheta_1(0|\tau)]^4 \left[\partial_z^{2}{G}(z)\right]^{N-2},
\end{equation}
where, since $\partial_z\vartheta_1(0|\tau)=2\pi\eta(\tau)^3$, the factor $[\partial_z\vartheta_1(0|\tau)]^4$ cancels out the denominator in eq. \eqref{partition function}.

Including the anti-holomorphic sector, the complete integral then reads ($z_2:={\rm Im}\,z$)
\begin{equation}
{\cal A}^{(N)}=\int_{\cal F} \frac{d^{2}{\tau}}{\tau_{2}^{5}} \int_{T^2} d^{2}{z} \, e^{-\frac{4\pi z_{2}^{2}}{\tau_{2}}(N{-}1)}\left|\frac{\vartheta_{1}(z|\tau)}{\partial_z\vartheta_1(0|\tau)}\right|^{4(N{-}1)} \left|{\partial_z^{2}{G}(z)}\right|^{2(N{-}2)},
\label{mass correction integral zero}
\end{equation}
where we have fixed one of the worldsheet insertions. We may write $\partial_z^2 G$ in the more compact form
\begin{align}
    \partial_z^2 G (z|\tau)=\tilde{e}_{1}(\tau)+f_{1}(\tau)\frac{\vartheta^2_2(z|\tau)}{\vartheta^2_1(z|\tau)},
\end{align}
with $\tilde{e}_{1}{:=}{-}4\pi i \partial_\tau {\log}\left[\sqrt{\tau_2}\vartheta_2(0|\tau)\right]$ and $f_{1}{:=}\frac{[\partial_z\vartheta_1(0|\tau)]^2}{\vartheta^2_2(0|\tau)}$. By the binomial identity, eq. \eqref{mass correction integral zero} becomes
\begin{flalign}
    {\cal A}^{(N)}&=\,\int_{\cal F}\frac{d^2\tau}{\tau_{2}^{5} \left|\partial_z\vartheta_1(0|\tau)\right|^{4(N-1)}}\int_{T^2} d^2z\, \,e^{-\frac{4\pi z_{2}^{2}}{\tau_{2}}(N-1)} \sum_{n=0}^{N-2}\binom{N{-}2}{n}\sum_{\bar{n}=0}^{N-2}\binom{N-2}{\bar{n}}\cdot&&\nonumber\\
    &\mathrlap{\quad\cdot\tilde{e}_{1}^{n}\bar{\tilde{e}}_{1}^{\bar{n}}f_{1}^{N-2-n}\bar{f}_{1}^{N-2-\bar{n}}\vartheta_{1}(z|\tau)^{2n+2}\overline{\vartheta_{1}(z|\tau)}^{2\bar{n}+2}\vartheta_{2}(z|\tau)^{2N-4-2n}\overline{\vartheta_{2}(z|\tau)}^{2N-4-2\bar{n}}.}&&
\label{mass correction integral}
\end{flalign}

We now closely examine the worldsheet integral
\begin{align}
    \mathcal{I}_{N_1,\overline{N}_1,N_2,\overline{N}_2}(\tau,\bar{\tau}) = {\int_0}^{\,1} dx {\int_0}^{\,1} dy \,e^{-4\pi i\tau_2 y^2(N-1)}&\vartheta_{1}(z|\tau)^{N_1}\overline{\vartheta_{1}(z|\tau)}^{\overline{N}_1}\cdot\nonumber\\
    &\cdot\vartheta_{2}(z|\tau)^{N_2}\overline{\vartheta_{2}(z|\tau)}^{\overline{N}_2},
\label{worldsheet integral}
\end{align}
with the level matching constraint $N_1{+}N_2{=}2(N{-}1){=}\overline{N}_1{+}\overline{N}_2$. Using the definitions of $\vartheta_1$ and $\vartheta_2$, the integral over $x$ gives a projector onto the allowed contributions and enables the recasting of the integral over $y$ in eq. \eqref{worldsheet integral} as a Gaussian integral, yielding ($q:=\exp(2i\pi\tau)$, $k_i,\,\hat{k}_i\in\mathbb{Z}\,\forall i$)
\begin{flalign}
    \mathcal{I}_{N_1,\overline{N}_1,N_2,\overline{N}_2}&=\,\frac{(-)^{\frac{N_1}{2}{+}\frac{\overline{N}_1}{2}}}{2\sqrt{\tau_{2}(N-1)}}{\sum_{h=1}^{2(N{-}1)}}e^{-\frac{\pi\tau_2 h^2}{N-1}}\left({\sum_{k_{1}}}{...}{\sum_{k_{N_1}}} {\sum_{\hat{k}_{1}}}{...} {\sum_{\hat{k}_{N_2}}}\right)\left({\sum_{\bar{k}_{1}}}{...}{\sum_{\bar{k}_{\overline{N}_1}}}{\sum_{\bar{\hat{k}}_{1}}}{...}{\sum_{\bar{\hat{k}}_{\overline{N}_2}}}\right){\cdot}&&\nonumber\\
    &\mathrlap{\cdot\delta_{\sum_{\ell=1}^{2(N-1)}k_{\ell},h}\delta_{\sum_{\ell=1}^{2(N-1)}\bar{k}_{\ell},h}q^{{1\over 2}\sum_{\ell=1}^{2(N-1)}k_{\ell}^{2}}\bar{q}^{{1\over 2}\sum_{\ell=1}^{2(N-1)}\bar{k}_{\ell}^{2}}\,e^{i\pi\left(\sum_{i=1}^{N_1}k_i - \sum_{\bar{i}=1}^{\overline{N}_1}\bar{k}_{\bar{i}}\right)},}&&\\
    \Rightarrow\quad \mathcal{A}^{(N)}=&\, \int_{\cal F}\frac{d^2\tau}{\tau_{2}^{4}\left|\partial_z \vartheta(0|\tau)\right|^{4(N-1)}} \sum_{n=0}^{N-2}\binom{N{-}2}{n}\sum_{\bar{n}=0}^{N-2}\binom{N{-}2}{\bar{n}} \tilde{e}_{1}(\tau)^{n}\overline{\tilde{e}_{1}(\tau)}^{\bar{n}}\nonumber\\
    &\,f_{1}(\tau)^{N-2-n}\overline{f_{1}(\tau)}^{N-2-\bar{n}}{\cal I}_{2n+2,\,2\bar{n}+2,\,2N-4-2n,\,2N-4-2\bar{n}}(\tau,\bar{\tau}).
\label{final expression for A}
\end{flalign}

The final step is to compute the integral over the fundamental domain $\mathcal{F}$, which diverges and needs to be regularized and renormalized. This is achieved by extending the $i\varepsilon$-prescription to string theory \cite{Witten:2013pra,Sen:2016ubf,Manschot:2024prc}, ensuring a consistent analytic continuation: when a long ``tube'' develops on the worldsheet, the prescription implements a passage from Euclidean to Lorentzian signature, thus regulating IR divergences. Reinstating the overall factors as in \cite{Marcus:1988vs}, we find
\begin{align}
    \delta{\cal M}^2_{\text{1-loop},\,N}=-\frac{g_s^2}{\sqrt{2(N{-}1)}\,(8\pi^2)^3}{\cal A}^{(N)},
    \label{mass shift}
\end{align}
where the real part is related to the mass shift and the imaginary part accounts for the decay width of the state. These corrections vanish for the massless level $N=1$, which is protected by gauge symmetry, while they are non-zero for massive states. The computation for the first massive level $N=2$ has been carried out in \cite{Marcus:1988vs,Stieberger:2023nol} and reproduced in \cite{Manschot:2024prc}. Our result
\begin{align}
    {\cal A}^{(N=2)}=(4.4687+i9.52381){\cdot}10^{-3}
\end{align}
agrees with the existing literature, up to an overall factor. Our scheme allows us to compute these corrections systematically, at arbitrary mass level. We focused on $N=3$ and $N=4$, yielding
\begin{align}
    {\cal A}^{(N=3)}=(1.699 + i1.708){\cdot}10^{-3}\quad\text{and}\quad {\cal A}^{(N=4)}=(7.975 + i6.242){\cdot}10^{-4}.
\end{align}
Our results suggest that the mass shifts decrease at large $N$. While too early to predict the suppression rate, our method enables this analysis for higher-mass states, which we leave for future work.

\section{Conclusions and outlook}
We presented a systematic method to compute one-loop mass corrections for NS–NS states in the leading Regge trajectory. Exploiting Jacobi $\vartheta_s$ functions and an extension of the $i\varepsilon$-prescription, we obtain a closed-form modular integrand at arbitrary mass level, enabling numerical evaluation.

Extending this approach to subleading trajectories and states with generic partitions of $N$ could shed light on the chaotic features of string scattering, though the relevant vertex operators become increasingly complex \cite{Bianchi:2010es, Basile:2024uxn}. As subleading trajectories admit multiple states with identical Lorentz quantum numbers at fixed mass, perturbative mixing and level repulsion may occur.

A meaningful assessment of complex behavior ultimately requires the study of heavier states, and the present results should be viewed as a first step in this direction.

\acknowledgments
We are grateful to P. Di Vecchia and E. Kiritsis for interesting discussions. LG also thanks M. Giachello, M. Bianchini and S. Tarquini for useful discussions and hospitality during 14YRM2025.

\bibliographystyle{JHEP}
\bibliography{stringbib}


\end{document}